\documentclass{aa}
\usepackage{graphicx}
\usepackage{txfonts}
\newcommand{\beq}{\begin{equation}}
\newcommand{\eeq}{\end{equation}}
\newcommand{\beqn}{\begin{eqnarray}}
\newcommand{\eeqn}{\end{eqnarray}}

\begin{document}
   \title{The generation of connected high and very high energy
   $\gamma$-rays and radio emission in active galactic nuclei}

  \author{Osmanov Z.
          }

   \offprints{Osmanov Z.}

   \institute{Centre for Theoretical Astrophysics, ITP, Ilia State University,
              0162 Tbilisi, Georgia\\
              \email{z.osmanov@astro-ge.org}
              }


  \abstract
  {}
   {We consider relativistic electrons in magnetospheric
   flows close to a supermassive black hole and study the mechanism of quasi-linear diffusion (QLD)
   to investigate the correlation between $\gamma$-ray
   and radio emission in active galactic nuclei.}
   {Moving in the nonuniform magnetic field the particles experience
   a force that is responsible for
   the conservation of the adiabatic invariant. This force, together with
   the radiation reaction force, tends to decrease the pitch angles. Contrary
   to this, the QLD attempts to increase the pitch
   angle, and this maintains the synchrotron emission regime. To
   examine
   the balance between the QLD and the aforementioned dissipative
   factors we investigate the quasi-stationary state by applying the kinetic
   equation.}
   {Considering the magnetospheric plasma close to the supermassive
   black hole, we examined the efficiency of the QLD for different
   parameters. By examining the cyclotron instability, we show that
   despite the efficient dissipative factors, the cyclotron modes
   excite transverse and longitudinal-transversal waves which leads to the QLD.
   We find that the QLD provides a connection
   of emission in the $\gamma$-ray and radio domains. We show that
   under favourable conditions the radio emission from
   $22$ MHz to $9$ GHz on the mpc scale is associated with $\gamma$-ray emission from
   $900$ GeV down to $9$ GeV on the same scale.}
   {}


   \maketitle
%


\section{Introduction}

During the last decades the interest in very high energy (VHE)
radiation of active galactic nuclei (AGNs) has considerably
increased because of the detailed data from the new telescopes
(MAGIC, HESS, EGRET, Fermi). Active galactic nuclei, and especially
blazars reveal a strong connection between high-energy $\gamma$-ray
and radio emission. Since the EGRET era, a direct connection between
radio and $\gamma$-ray flux density has been intensively discussed.
In particular, Bloom (2008) performed a statistical analysis of the
broadband properties of EGRET blazars and showed a strong positive
correlation between the total radio luminosity at $8.4$ GHz and the
$\gamma$-ray luminosity in the EGRET band ($L_{\gamma}\sim
L_{rad}^{0.77}$). The comparison of the radio flux density at $8.4$
GHz and the $\gamma$-ray ($>100$ MeV) photon flux measured by the
Fermi $\gamma$-ray Space Telescope has been presented by Giroletti
et al. (2010) and the Spearman's rank correlation coefficient was
estimated to be $0.57$. Kovalev et al. (2009) found a positive
correlation between the pc-scale radio flux at $15$ GHz and the
$\gamma$-ray photon flux in a sub-band of the Fermi/LAT detector
($100$ MeV-$1$ GeV), and a time separation of a few months between
$\gamma$-ray and radio flares.

Most commonly the VHE emission of AGNs is explained in terms of the
inverse Compton upscattering (\cite{blan}) and curvature radiation
(\cite{g96,tg05}). The major reason for this is that if the magnetic
field is strong, the synchrotron cooling timescale is very short and
relativistic particles very soon transit to their ground Landau
level, which prevents the subsequent emission process
(\cite{difus3}).

Recently, it has been shown that the quasi-linear diffusion (QLD)
may create necessary conditions in the magnetospheres of AGNs and
pulsars for avoiding the synchrotron emission damping despite the
strong magnetic field. According to the model of Kazbegi et al.
(1992), the cyclotron instability appears in the magnetospheres of
pulsars. These unstable modes lead to the feedback of the excited
waves on particles by means of the process of the diffusion. The
pitch angles are no longer negligible and the emission process
continues (e.g. \cite{machus1,lomin,malmach,nino}). This model was
applied to the Crab pulsar (\cite{difus,difus1}) to explain the
recent results from the MAGIC Cherenkov telescope between 2007
October and 2008 February (\cite{magic}).

In the context of works presented by Bloom (2008) and Giroletti et
al. (2010) (where the observationally evident strong correlation of
high-energy $\gamma$-rays and low-frequency radiation is discussed),
it is of fundamental importance that according to the QLD, two
different radiation domains are excited. As was studied by Osmanov
\& Machabeli (2010) and Osmanov (2010), the cyclotron resonance
efficiently generates transverse momenta, which leads to non-zero
pitch angles and the subsequent synchrotron radiation. This means
that a connection between two radiation domains is naturally
introduced into the quasi-linear diffusion process. In particular,
examining the QLD in AGNs, Osmanov \& Machabeli (2010) studied the
generation of $X$-rays associated with the radio emission, and the
work of Osmanov (2010) was related to high-energy emission
associated with submillimeter/infrared radiation.

In this paper we study the production of high-energy radiation in
the GeV-TeV interval in the magnetospheres of supermassive black
holes. The paper is organized as follows. In Section 2 we present
the model, in Sect. 3 we apply the mechanism to AGNs and in Sect. 4
we summarize our results.

\section{Theory} \label{sec:consid}

In general, Lorentz factors of magnetospheric plasma particles in
AGNs\footnote{ The magnetosphere of AGNs with their typical
lengthscales $10^{14-15}$ cm is a region around the supermassive
black hole, where the magnetic field ($10^4$ G - close to the black
hole and $10-100$ G - in the light cylinder zone) is dominant.} lie
in a broad interval ranging from $\sim 1$ up to $\sim 10^8$
(\cite{osm7,ra08}), therefore, for simplicity one can consider the
magnetosphere to be composed of the electron-positron plasma
component with relatively low Lorentz factors, $\gamma_p$, and
highly relativistic electrons, the so-called beam component
electrons with the Lorentz factor $\gamma_b$
($\gamma_b\gg\gamma_p$).

In this section we consider the QLD mechanism and derive an average
value of the pitch angles driven by the diffusion. In general the
process of energy dissipation that takes place in strong magnetic
fields is dynamically controlled by two forces: one is the
synchrotron radiative force (\cite{landau})
\begin{equation}\label{f}
F_{\perp} = -\alpha\psi(1 + \gamma_b^2\psi^2),\;\;\;\;\;F_{_{\|}} =
-\alpha\gamma_b^2\psi^2,
\end{equation}
where $\alpha = 2e^2\omega_B^2/(3c^2)$, $\psi$ is the pitch angle,
$\omega_B\equiv eB/mc$ is the cyclotron frequency, $B$ is the
magnetic induction, $c$ is the speed of light, $e$ and $m$ are
electron's charge and the rest mass, respectively; the other is the
force ${\bf G}$, responsible for the conservation of the adiabatic
invariant, $I = 3cp_{\perp}^2/2eB$ (\cite{landau})
\begin{equation}\label{g}
G_{\perp} = -\frac{mc^2}{\rho}\gamma_b\psi,\;\;\;\;\;G_{_{\|}} =
\frac{mc^2}{\rho}\gamma_b\psi^2,
\end{equation}
where $\rho$ is the curvature radius of the magnetic field lines. In
general, in the magnetospheres of AGNs two effects take place: the
energy dissipation provoked by the forces, ${\bf F, G}$, which lead
to the decrease of the pitch angles, the QLD, which opposes the
first effect. The dynamical process saturates when the effects of
the above-mentioned forces are balanced by the diffusion, which
leads to non-zero pitch angles and the subsequent radiation.

As discussed by Machabeli \& Usov (1979), Machabeli \& Osmanov
(2010), Osmanov \& Machabeli (2010), for an activation of the QLD,
the cyclotron modes must be excited, which leads to the diffusion in
terms of feedback of the cyclotron waves on particles. On the other
hand, according to the work of Kazbegi et al. (1992), the anomalous
Doppler effect can efficiently induce unstable cyclotron waves with
the corresponding frequency (\cite{machus1,difus3})
\begin{equation}\label{om1}
\nu\approx \frac{\omega_B}{2\pi\delta\cdot\gamma_b},
\end{equation}
where $\delta = \omega_p^2/(4\omega_B^2\gamma_p^3)$, $\omega_p
\equiv \sqrt{4\pi n_pe^2/m}$ is the plasma frequency and $n_p$ is
the plasma density. Here we assume that energy in plasma is
uniformly distributed, therefore, the plasma density can be
approximated to be $n_b\gamma_b/\gamma_p$, where $n_b$ is the beam
density (\cite{difus}). Although the aforementioned expression is a
direct consequence of a resonance character of the cyclotron
instability, in exciting waves all resonance particles (with broad
energy spectra) participate, therefore the range of spectral
frequencies is wide and the waves are not characterized by a
monochromatic signature. In the framework of the present approach,
one has two simultaneously generated connected VHE $\gamma$-rays and
low-frequency cyclotron modes. This result seems to be promising in
the context of recent investigations, because it shows the strong
connection between the $\gamma$-ray and radio fluxes
(\cite{bloom,giro}).

A similar problem was studied by Osmanov \& Machabeli (2010) who
examined the physical regime $|G_{\perp}|\gg |F_{\perp}|$ and
$|G_{_\parallel}|\ll |F_{_\parallel}|$ and studied the production of
X-rays ($0.13$ keV-$100$ keV) connected with the radio emission
($0.04$ MHz-$35$ MHz). A different regime, $|G_{\perp}|\ll
|F_{\perp}|$ and $|G_{_\parallel}|\ll |F_{_\parallel}|$, was
examined by Osmanov (2010). This regime shows the possibility of
production of MeV-GeV $\gamma$-rays, which are strongly correlated
with submillimeter/infrared radiation domains. In the present paper
we examine the same regime, but for different parameters. Unlike the
previous work, where the QLD was studied in the light cylinder area
where the magnetic field is of the order of $\sim 300$G, in the
present paper the physics of the QLD is examined for a region close
to the supermassive black hole event horizon where $B\sim 10^4$G
(see \cite{paradig}). Another difference is the fact that here we
study the cyclotron waves generated by plasma particles with Lorentz
factors, $\gamma_p\sim 2-4$, whereas values of $\gamma_p$ considered
by Osmanov (2010) were higher by two orders of magnitude. As we will
see, the aforementioned differences significantly change the
results. By comparing the transverse and longitudinal components of
the forces, one obtains
\begin{equation}\label{gfpp}
\frac{G_{\perp}}{F_{\perp}}\approx 5\times
10^{-8}\times\left(\frac{10^8}{\gamma_b}\right)\times\left(\frac{10^4
G}{B}\right)^2\times\left(\frac{10^{-2}
rad}{\psi}\right)^2\times\left(\frac{R_g}{\rho}\right),
\end{equation}
\begin{equation}\label{gfpr}
\frac{|G_{_\parallel}|}{|F_{_\parallel}|}\approx 5\times
10^{-12}\times\left(\frac{10^8}{\gamma_b}\right)\times\left(\frac{10^4
G}{B}\right)^2\times\left(\frac{R_g}{\rho}\right).
\end{equation}
Evidently, the relations $|G_{\perp}|\ll |F_{\perp}|$ and
$|G_{_\parallel}|\ll |F_{_\parallel}|$ are satisfied if $\psi\gg
2.2\times 10^{-6}$ rad if we consider the physical parameters
$\gamma_b\sim 10^8$, $B\sim 10^4$ G, $R_g\sim\rho$.

By assuming a quasi-stationary scenario ($\partial/\partial t = 0$),
the corresponding kinetic equation will be different from that of
Machabeli \& Osmanov (2010)
$$\frac{1}{mc\gamma_b\psi}\frac{\partial}{\partial\psi}
\left(\psi F_{_\perp}f\right) +
\frac{1}{mc}\frac{\partial}{\partial\gamma_b}\left(F_{_\parallel}
f\right) + \upsilon\frac{\partial f}{\partial r}=$$

$$=\frac{1}{m^2c^2\gamma_b^2\psi}\frac{\partial}{\partial\psi}
\left(\psi D_{_{\perp\perp}}\frac{\partial
f}{\partial\psi}\right)+\frac{1}{mc\psi}\frac{\partial}{\partial\psi}
\left(\psi^2 D_{_{\perp\parallel}}\frac{\partial
f}{\partial\gamma_b}\right)+$$

\begin{equation}\label{kinet}
\;\;\;\;\;\;\;\;+\frac{1}{mc}\frac{\partial}{\partial\gamma_b}
\left(\psi D_{_{\perp\parallel}}\frac{\partial
f}{\partial\psi}\right),
\end{equation}
where $f = f(\psi,\gamma_b)$ is the distribution function of
particles,
\begin{equation}\label{dif}
D_{\perp\perp}\approx \frac{\pi^2
e^2}{m^2c^3}\frac{\delta}{\gamma_b^2}|E_k|^2,\;\;\;\;\;\;
D_{\perp_{\parallel}}\approx -\frac{\pi e^2}{4mc^2\gamma_b}|E_k|^2,
\end{equation}
are the diffusion coefficients and $|E_k|^2$ is the energy density
per unit wavelength.

For estimating $|E_k|^2$, we assume that half of the plasma energy
density, $mc^2n_b\gamma_b/2$, converts to the energy density of the
waves $|E_k|^2k$, then an expression of $|E_k|^2$ writes as follows
\begin{equation}\label{ek2}
|E_k|^2 = \frac{mc^3n_b\gamma_b} {2\omega}.
\end{equation}
For solving the aforementioned kinetic equation with respect to the
pitch angles we separate the variables by expressing the
distribution function as $\chi(\psi)f(\gamma_b)$. Then the equation
governing the distribution by pitch angles is given by
\begin{equation}\label{kinet1}
\frac{\partial}{\partial\psi} \left(\psi
F_{_\perp}f\right)=\frac{1}{mc\gamma_b}\frac{\partial}{\partial\psi}
\left(\psi D_{_{\perp\perp}}\frac{\partial f}{\partial\psi}\right),
\end{equation}
and the corresponding solution writes as
\begin{equation}\label{chi} \chi(\psi) = Ce^{-A\psi^4},
\end{equation}
where
\begin{equation}\label{A}
A\equiv \frac{\alpha mc\gamma_b^3}{4D_{_{\perp\perp}}}.
\end{equation}

For simplicity we use the average value of the pitch angle, which
after taking Eqs. (\ref{chi},\ref{A}) into account leads to
\begin{equation}\label{pitch}
\bar{\psi}
 = \frac{\int_{0}^{\infty}\psi \chi(\psi)d\psi}{\int_{0}^{\infty}\chi(\psi)d\psi}
\approx \frac{0.5}{\sqrt[4]{A}}.
\end{equation}

Clearly, despite very efficient synchrotron losses the quasi-linear
diffusion works against the dissipative forces and maintains
non-zero pitch angles, which prevent the synchrotron mechanism from
damping. Under these conditions, the relativistic particles with
Lorentz factors, $\gamma$, radiate in the synchrotron regime and
emit photons with energies (\cite{Lightman})
\begin{equation}
\label{eps} \epsilon_{eV}\approx 1.2\times 10^{-8}B\gamma^2\sin\psi.
\end{equation}

\section{Results}
\begin{figure}
  \resizebox{\hsize}{!}{\includegraphics[angle=0]{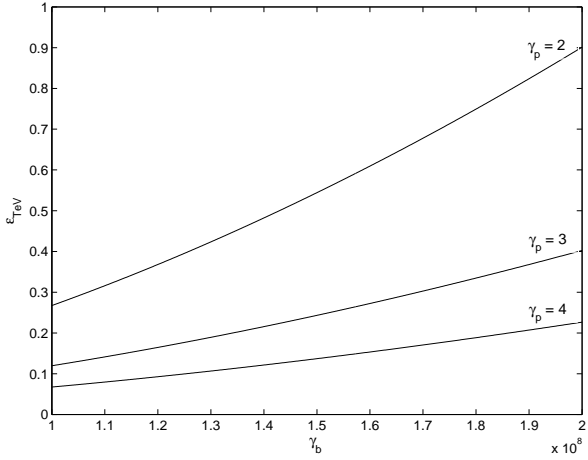}}
  \caption{Behaviour of $\epsilon_{TeV}$ with respect to $\gamma_b$.
The set of parameters is $M_{9} = 1$, $B = 10^4$ G, $n_b = 20$
 cm$^{-3}$ and $\gamma_p\in\{2;3;4\}$.}\label{fig1}
\end{figure}

According to the observational evidence, AGNs are characterized by
emission in the VHE domain, which cannot be explained in the
framework of thermal processes. This can also be explained through
the existence of ultra-relativistic particles in the magnetosphere.
In general, there are several acceleration mechanisms that might
provide very high Lorentz factors. We show that processes such as
Fermi-type acceleration (\cite{cw99}) and acceleration owing the the
black hole dynamo mechanism (\cite{levins}) can provide Lorentz
factors of the order of $\sim 10^{7-9}$ close to the supermassive
black holes, i.e. at $r\sim R_g$, where $R_g\equiv
2GM_{BH}/c^2\approx 3M_9\times 10^{14}$ cm is the gravitational
radius of the black hole, $M_9\equiv M/(10^9M_{\odot})$ and
$M_{\odot}$ is the solar mass.

\begin{figure}
  \resizebox{\hsize}{!}{\includegraphics[angle=0]{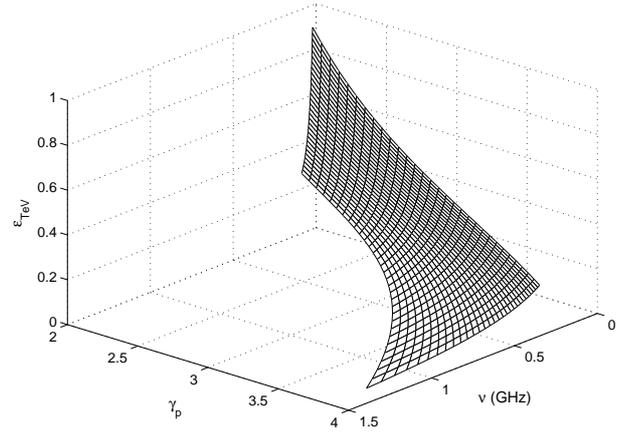}}
  \caption{Behaviour of $\epsilon_{TeV}$ versus $\nu$ and $\gamma_p$.
The set of parameters is $M_{9} = 1$ and $B = 10^4$ G, $n_b = 20$
 cm$^{-3}$.}\label{fig2}
\end{figure}
Osmanov \& Machabeli (2010) have shown that the synchrotron cooling
timescale for ultra-relativistic electrons in the magnetospheres of
AGNs close to the black hole is of the order of $\sim
10^{-5}$s-$10^{-4}$s. To estimate the efficiency of the energy
losses it is reasonable to calculate the kinematic timescale
$t_{kin}\sim R_g/c$. Considering the supermassive black hole with
$M_9=1$, one can see that $t_{kin}\sim 10^4$s, which by many orders
of magnitude exceeds the dissipation timescale, therefore, the
cooling process is extremely efficient and the electrons stop
emitting very soon.

Anomalous Doppler effect significantly changes the overall radiative
pattern of the system.  Equation (\ref{om1}) clearly shows that the
mentioned effect efficiently induces unstable cyclotron waves with
the frequency (\cite{difus3})
\begin{equation}
\label{om} \nu\approx 8.7\times
10^7\left(\frac{\gamma_p}{2}\right)^4\times\left(\frac{10^8}
{\gamma_b}\right)^2\times\left(\frac{B}{10^4~G}\right)^3
\times\left(\frac{20~cm^{-3}}{n_b}\right)Hz,
\end{equation}
which in turn lead to the feedback of these modes to the particles
and switch on the QLD, which generates the non-vanishing pitch
angles and the subsequent synchrotron process.

As a first example, we examine a supermassive black hole with $M_9
=1$ and the typical value of the magnetic induction for a nearby
zone of the black hole $B = 10^4$ G. Since we assume that energy in
plasma is uniformly distributed, the value of plasma density
$n_b\gamma_b/\gamma_p$, for $n_b = 20$ cm$^{-3}$ and $B = 10^4$ G is
of the order of $10^9$ cm$^{-3}$, which is compatible with
observations of narrow and broad line regions and indicates that the
plasma density in the environment of AGNs lies in the interval $\sim
(10^5-10^{14})$ cm$^{-3}$ (\cite{xu,arav}). We consider the beam
component's Lorentz factors to be of the order of $10^8$ and the
plasma component with the Lorentz factor, $\gamma_p = 2$. From Eq.
(\ref{om}) one can show that the cyclotron frequency of the order of
$87$ MHz is excited, which in turn, by means of the QLD, leads to
the average value of the pitch angle $\sim 0.2$ rad (see Eqs.
[\ref{A},\ref{pitch}]) and the subsequent synchrotron emission in
the VHE domain $\sim 240$ GeV [see Eq. (\ref{eps})].

In Fig. \ref{fig1} we show the dependence of $\epsilon_{_{TeV}}$ on
$\gamma_b$. The set of parameters is $M_{9} = 1$, $B = 10^4$ G, $n_b
= 20$ cm$^{-3}$ and $\gamma_p\in\{2;3;4\}$. Clearly, physically
realistic parameters might guarantee the synchrotron emission from
$\sim 70$ GeV ($\gamma_b = 10^8$, $\gamma_p = 4$) to $0.9$ TeV
($\gamma_b = 2\times 10^8$, $\gamma_p = 2$). These plots show a
continuously increasing character of $\epsilon_{_{TeV}}$ with
respect to $\gamma_b$, which is a natural result of the fact that
more energetic beam particles radiate photons with higher energies.
Unlike the beam Lorentz factors, the behaviour of
$\epsilon_{_{TeV}}$ with respect to $\gamma_p$ is a continuously
decreasing function. This can be seen from Eqs.
(\ref{A},\ref{pitch}): $\bar{\psi}\sim\sqrt[4]{D_{\perp\perp}}$,
which by combining with Eqs. (\ref{dif},\ref{ek2}) reduces to
$\bar{\psi}\sim 1/\gamma_p^2$. Therefore, by considering higher
values of $\gamma_p$, the resulting synchrotron emission energy will
be lower.

\begin{figure}
  \resizebox{\hsize}{!}{\includegraphics[angle=0]{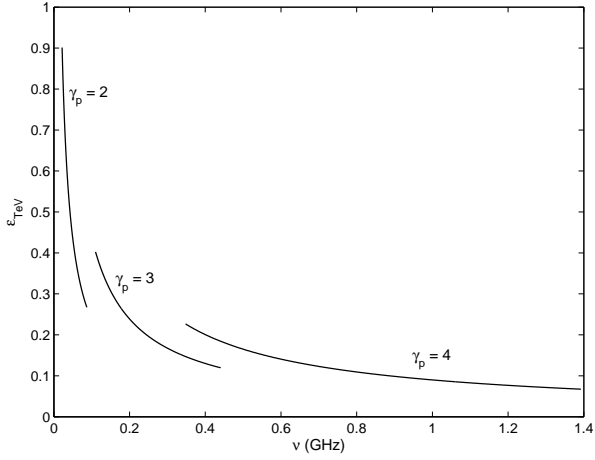}}
  \caption{Behaviour of $\epsilon_{TeV}$ versus $\nu$.
The set of parameters is $M_{9} = 1$ and $B = 10^4$ G, $n_b = 20$
 cm$^{-3}$ and $\gamma_p = \{2;3;4\}$. Evidentlythe curves are
truncated, which is a direct result of the fact that the curves are
one-dimensional sections of a two-dimensional surface.}\label{fig3}
\end{figure}

In Fig. \ref{fig2} we show the behaviour of  $\epsilon_{TeV}$ with
respect to $\nu$ and $\gamma_p$. The set of parameters is $M_{9} =
1$ and $B = 10^4$ G, $n_b = 20$ cm$^{-3}$. The 2D surface shows that
for lower values of the plasma Lorentz factors, the synchrotron
emission energy is higher and the corresponding cyclotron frequency
relatively is lower. Fig.\ref{fig3} shows the dependance of
$\epsilon_{_{TeV}}$ on the cyclotron radio frequency for three
different values of plasma Lorentz factors $\gamma_p=\{2;3;4\}$.
Evidently, the VHE $\gamma$-rays in the TeV domain are connected
with the radio emission at $\sim 87$ MHz, and the higher frequency
radio emission $\sim 1.4$ GHz is connected with the lower energy
($\sim 70$ GeV) $\gamma$-ray synchrotron emission.

\begin{figure}
  \resizebox{\hsize}{!}{\includegraphics[angle=0]{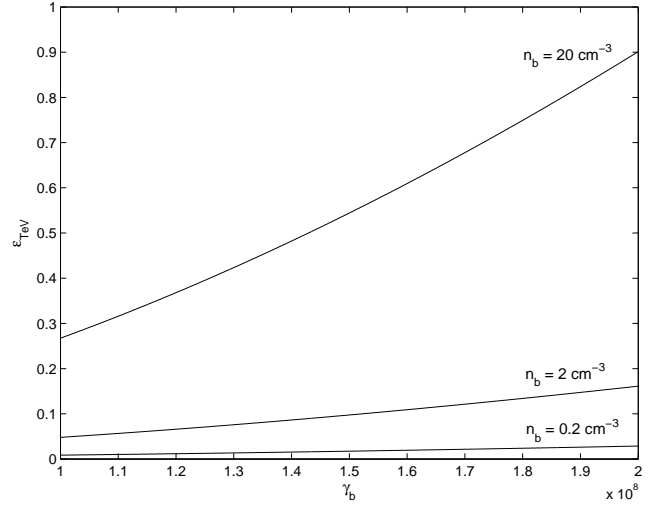}}
  \caption{Behaviour of $\epsilon_{TeV}$ versus $\gamma_b$.
The set of parameters is $M_{9} = 1$ and $B = 10^4$ G, $\gamma_p =
2$ and $n_b=\{0.2;2;20\}$ cm$^{-3}$.}\label{fig4}
\end{figure}

One of the major magnetospheric parameters is the density of the
beam component. In Fig. \ref{fig4} we show the behaviour of
$\epsilon_{_{TeV}}$ versus $n_b$. The set of parameters is $M_{9} =
1$ and $B = 10^4$ G, $\gamma_p = 2$ and $n_b=\{0.2;2;20\}$
cm$^{-3}$. Obviously, the higher the beam density, the higher the
$\gamma$-ray synchrotron radiation energy. As we  already mentioned,
the average value of the pitch angle is proportional to
$\sqrt[4]{D_{\perp\perp}}$, which by combining with
$D_{\perp\perp}\sim n_b^3$ [see Eqs. (\ref{dif},\ref{ek2})] confirms
the results shown in the plots. For the considered interval of $n_b$
the photon energy varies in the range, $8.5$ GeV-$0.9$ TeV.

\begin{figure}
  \resizebox{\hsize}{!}{\includegraphics[angle=0]{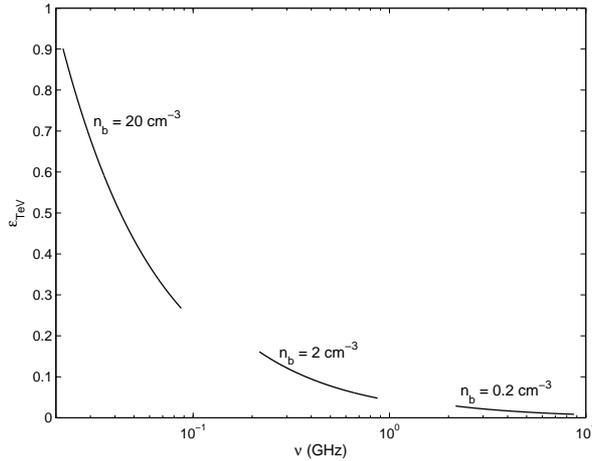}}
  \caption{Behaviour of $\epsilon_{TeV}$ versus $\nu$.
The set of parameters is $M_{9} = 1$ and $B = 10^4$ G, $\gamma_p =
2$ and $n_b=\{0.2;2;20\}$ cm$^{-3}$. The curves are truncated for
the same reason as in Fig. \ref{fig3}.}\label{fig5}
\end{figure}

Fig. \ref{fig5} presents the dependance of $\epsilon_{_{TeV}}$ on
$\nu$. From the plots we see that the TeV emission is connected to
the radio band $\sim 22$ MHz and the high-energy emission, $8.5$ GeV
is connected to $\sim 8.7$ GHz radio domain.
The present results are different from those of Osmanov \& Machabeli
(2010) and Osmanov (2010), where a different parametric space, has
been considered. In particular, Osmanov \& Machabeli (2010) examined
$M_9 = 1$, $\gamma_p=2$, $\gamma_b\sim 10^{6-9}$, $n_b=2000$
cm$^{-3}$ and Osmanov (2010) considered $M_9 = 1$,
$\gamma_p\in\{200;250;300\}$, $\gamma_b\sim 10^8$ and
$n_b=\{5;10;15\}$ cm$^{-3}$. 

According to the quasi-linear approach, the efficiency of the QLD
depends on the efficiency of the cyclotron instability. For this
purpose it is essential to consider the growth rate of the
instability. Kazbegi et al. (1992) showed that the increment,
$\Gamma$, of the modes is given by
\begin{equation}\label{inc1}
\Gamma = \pi \frac{\omega_b^2}{\omega\gamma_p} \;\;\;\ if \;\;\;\
\frac{1}{2}\frac{u_x^2}{c^2}\ll\delta
\end{equation}
and
\begin{equation}\label{inc2}
\Gamma = \pi
\frac{\omega_b^2}{2\omega\gamma_p}\frac{u_x^2}{\delta\cdot c^2}
\;\;\;\ if \;\;\;\ \frac{1}{2}\frac{u_x^2}{c^2}\gg\delta,
\end{equation}
where $\omega_b\equiv\sqrt{4\pi n_b e^2/m}$ is the plasma frequency
of resonant (beam) electrons. For the parameters $n_b = 20$
cm$^{-3}$, $\gamma_p = 2$ (see Fig. \ref{fig1}), by assuming
$u_x\sim c$ and $\rho\sim R_g$, one can show that $u_x^2/(2c^2)\ll
1$, and the growth rate is defined by Eq. (\ref{inc1}). Therefore,
the corresponding cyclotron timescale, $t_{cyc}=1/\Gamma$, is of the
order of $10^{-3}$ s, which is less by many orders of magnitude than
the kinematic timescale, $t_{kin}\sim R_g/c\sim 10^{4}$ s. The same
relation between the timescales is valid for the rest of the
considered cases; therefore, the studied mechanism is extremely
efficient.

The interest in the QLD is twofold: on the one hand, the
quasi-linear diffusion is an efficient mechanism that drives the
synchrotron process despite very short synchrotron cooling
timescales. On the other hand, as we have seen, the QLD provides the
possibility of generating connected VHE $\gamma$-rays and radio
emission on mpc scales. In particular, Bloom (2008) has found a high
correlation of $\gamma$-ray luminosity and radio luminosity. The
work of Giroletti et. al (2010) is particularly interesting, because
they show that radio emission at $8.4$ GHz is connected to the VHE
radiation ($>100$ MeV). As we show in Fig. \ref{fig5}, the radio
emission of the order of $8.4$ GHz is associated with $\gamma$-rays
in the GeV band, therefore, the study of the QLD is very promising.


\section{Summary}\label{sec:summary}

The main aspects of the present work can be summarized as follows:
\begin{enumerate}

      \item We studied the role of the quasi-linear diffusion
      with magnetospheric plasma particles in AGNs. For this purpose
      we considered the kinetic
      equation that describes the balance between the synchrotron
      reaction force and the diffusion.

      \item We  examined our model for different values of
      physical parameters. The QLD was studied versus the Lorentz
      factors of plasma and beam components respectively and versus
      the density of the beam component. We found that the higher
      the beam Lorentz factor, the higher the synchrotron emission
      energy and moreover, that the higher the plasma Lorentz factor,
      the lower the radio frequency. We also showed that
      lower beam densities lead to lower energies in the
      $\gamma$-ray synchrotron, and to higher energies in the radio domain.

      \item We also found that under favourable conditions the
      cyclotron instability generates radio emission from
      $\sim 22$ MHz to $\sim 9$ GHz and the QLD provides
      synchrotron energies in the interval $900$ down to $9$ GeV. These
      results show the existence of connected radio and $\gamma$-ray emissions on mpc scales.

      \end{enumerate}

The QLD is a mechanism able to explain a connection between radio
emission and VHE $\gamma$-rays. On the other hand, a complete study
requires an investigation of the spectral pattern provided by the
mentioned mechanism. According to the standard theory of the
synchrotron emission (\cite{bekefi,ginz}), it is assumed that along
the line of sight the magnetic field is chaotic; correspondingly,
the pitch angles vary in a broad interval (from $0$ to $\pi$) and
the particles have uniformly distributed pitch angles. Contrary to
this scenario, in the framework of the present model the particles
are nontrivially distributed (see Eq. \ref{chi}), therefore, the
spectral picture will differ from those of Bekefi \& Barrett (1977),
Ginzburg (1981). Another problem that has to be addressed is the
variability of the corresponding VHE radiation. As the investigation
shows, for the QLD to be efficient, two major requirements have to
be fulfilled: the particles must be highly relativistic and the
timescale of the cyclotron instability should be short enough.
Therefore, the study of the variability implies the corresponding
study of the acceleration timescale and the aforementioned
instability timescale. We will investigate this problem in future
studies.

\section*{Acknowledgments}
The research was supported by the Georgian National Science
Foundation grant GNSF/ST07/4-193. I also thank an anonymous referee
for helpful suggestions.

\end{document}